\documentclass{PoS}

\usepackage[utf8x]{inputenc}
\usepackage[pdftex]{graphicx}
\usepackage{stmaryrd}
\usepackage{slashed}
\usepackage{textcomp}
\usepackage{subfigure}
\usepackage{amsmath}
\usepackage{amssymb}
\newcommand{\I}{\ensuremath{\mathrm{i}\hspace{1pt}}}
\newcommand{\arxiv}[1]{arXiv:\,\href{http://arxiv.org/abs/#1}{{\tt #1}}}

\title{First studies of the phase diagram of $\mathcal{N}=1$ supersymmetric Yang-Mills theory}

\ShortTitle{First studies of the phase diagram of $\mathcal{N}=1$ supersymmetric Yang-Mills theory}

\author{
G.~Bergner\\
Universit\"at Frankfurt, Institut f\"ur Theoretische Physik\\
Max-von-Laue-Str.~1, D-60438 Frankfurt am Main, Germany\\
E-mail: bergner@th.physik.uni-frankfurt.de}
\author{P.~Giudice, G.~M\"unster, \speaker{S.~Piemonte}, D.~Sandbrink\\
Universit\"at M\"unster, Institut f\"ur Theoretische Physik\\
Wilhelm-Klemm-Str.~9, D-48149 M\"unster, Germany\\
E-mail: munsteg, p.giudice, spiemonte, dirk.sandbrink@uni-muenster.de
}

\abstract{The behavior of supersymmetric theories at finite temperatures differs from that of other theories in certain aspects. Due to the different thermal statistics of bosons and fermions, supersymmetry is explicitly broken for any non-zero value of the temperature. We study $\mathcal{N}=1$ supersymmetric Yang-Mills theory on the lattice at finite temperatures. This model is the simplest supersymmetric extension of the pure gauge sector of QCD, describing the interactions between gluons and their fermionic superpartners, the gluinos. At zero temperature the theory confines like QCD, and chiral symmetry is spontaneously broken. At high temperatures, deconfinement and chiral symmetry restoration are expected to take place, but it is not known whether these two phase transitions coincide or not. First results on this topic, obtained in numerical simulations on the lattice, will be presented and discussed.}

\FullConference{The 32nd International Symposium on Lattice Field Theory\\
                 23-28 June, 2014\\
                 Columbia University New York, NY}

\begin{document}

The fundamental symmetries of $\mathcal{N}=1$ super Yang-Mills (SYM) theory have two opposite thermodynamical behaviors. Scale and chiral invariance are expected to be restored at sufficient high temperatures, but on the contrary supersymmetry itself is explicitly broken unless the temperature is zero \cite{Girardello:1980vv}. Supersymmetry is therefore expected to play only a little role in the thermodynamics of SYM models and results can be qualitatively extended to strongly interacting theories without supersymmetry.

The $\mathcal{N}=1$ super Yang-Mills theory shows confinement at low temperatures and chiral symmetry is spontaneously broken, like in QCD \cite{Amati:1988ft,Bergner:2012rv,Bergner:2013nwa}. However in $\mathcal{N}=1$ SYM both phenomena are realized exactly and define two distinct phases separated by a critical point: the relation between deconfinement and chiral symmetry restoration can be studied exactly, because in this theory there are exact order parameters for the transitions, see the following sections. In QCD chiral and deconfinement phase transition occur approximately at the same temperature, but it is unclear whether this fact is a coincidence due to the lack of an exact order parameter \cite{Aoki:2006br,Borsanyi:2010bp}. Numerical simulations of $\mathcal{N}=1$ SYM can provide thus a better understanding of the thermodynamics of strong interactions and can test the correctness of the effective models that have been proposed so far, see for instance Ref.~\cite{Aharony:2006da,Davies1999,Lacroix2014}.

In this contribution we present the first lattice results about the phase diagram of $\mathcal{N} = 1$ supersymmetric Yang-Mills theory with gauge group SU(2) at finite temperatures. The detailed discussion regarding the main simulations have been published in Ref.~\cite{Bergner2014}.

\section{The model}

The $\mathcal{N}=1$ supersymmetric Yang-Mills (SYM) theory is a SUSY model that describes strong interactions between gluons and ``gluinos''. The gluino is a spin-\textonehalf~Majorana fermion in the adjoint representation of the gauge group and it is the superpartner of the gluon. The action on-shell $S$ reads in the continuum
\begin{equation}
 S = \int d^4 x \left( -\frac{1}{4} \textrm{Tr}( F_{\mu\nu}F^{\mu\nu} ) + \frac{1}{2} \bar{\lambda}(x)(i\slashed{D})\lambda(x) \right)\,,
\end{equation}
and it is equal to the action of one flavor QCD if the gluino $\lambda(x)$ is replaced by a Dirac fermion field in the fundamental representation. 

In our simulations the gauge part of $S$ is discretized on the lattice with the Symanzik tree-level improved action $S_{WS}$, while $\slashed{D}$ is discretized using the Dirac-Wilson operator $D_W$. A gluino mass term is introduced to fine tune the lattice theory to a consistent continuum limit where the renormalized gluino mass vanishes and supersymmetry is dynamically restored \cite{Curci:1986sm,Suzuki:2012pc}.

The gluino fermion field can be integrated out in the partition function as in QCD but the result is the Pfaffian of the Dirac-Wilson operator
\begin{equation}
 \mathcal{Z} = \int D U_\mu(x) \textrm{Pf}(C D_W) \exp{\left\{-\beta S_{WS} \right\}}\,.
\end{equation}
The Pfaffian can be positive or negative unlike the determinant of this theory and therefore the integrand above
cannot be interpreted as a probability weight. The Pfaffian is equal up to its sign to the square root of the determinant: configurations are generated by the Hybrid Monte Carlo algorithm accordingly to the distribution
\begin{equation}
 \mathcal{Z}' = \int D U_\mu(x) \sqrt{\textrm{det}(C D_W)} \exp{\left\{-\beta S_{WS} \right\}}\,,
\end{equation}
while the sign of $\textrm{Pf}(C D_W)$ is inserted as reweighting factor in Monte Carlo simulations. A sign problem will appear if $\langle \textrm{sign}(\textrm{Pf}(C D_W))\rangle = 0$. The sign of the Pfaffian has been measured and it is normally always positive if the gluino mass is sufficiently far from the chiral limit.

\section{The deconfinement phase transition}

\begin{figure}[t]
\centering
\subfigure[$\langle| P_L |\rangle$]{\includegraphics[width=0.49\textwidth]{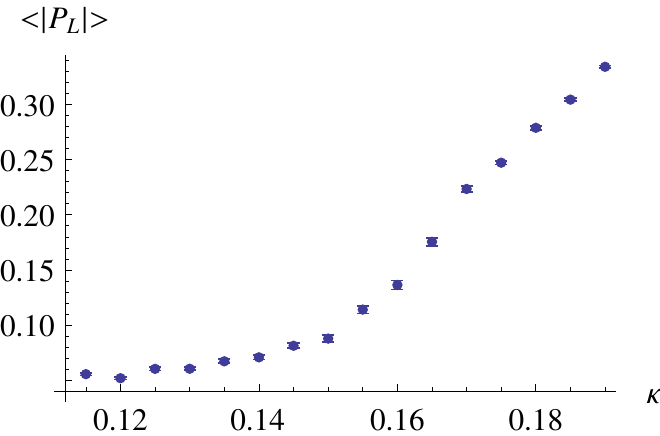}\label{Polyakov12_1650b.pdf}}
\subfigure[$\chi_P$]{\includegraphics[width=0.49\textwidth]{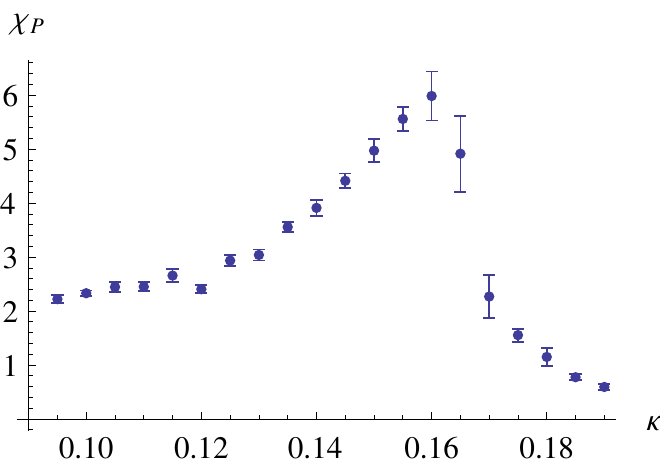}\label{SusceptibilityPolyakov12_1650b.pdf}}
\caption{Expectation value of the Polyakov loop and its susceptibility as a function of $\kappa=\frac{1}{2(m_0+4)}$ on a lattice $12^3\times4$ with $\beta = 1.65$.}
\end{figure}

The Polyakov loop is an order parameter for the deconfinement phase transition. Center symmetry is preserved in the fermion action of our model, unlike in QCD, since the Dirac-Wilson operator acts on the gluino field in the adjoint representation of the gauge group. The Polyakov loop is thus an exact order parameter for any value of the gluino mass.

The bare parameter space has been explored at finite temperatures varying the bare gluino mass $m_0$ at several fixed values of the gauge coupling $g$ to search the point where the deconfinement phase transition occurs. For example, in Fig.~\ref{Polyakov12_1650b.pdf} the expectation value of the modulus of the Polyakov loop $\langle| P_L |\rangle$ is shown as a function of $\kappa=1/(2m_0+8)$ for a fixed $\beta = (2N_c)/g^2 = 1.65$ on a lattice $12^3\times4$; $\langle| P_L |\rangle$ acquires a non-zero value at $\kappa\approx 0.155$. The susceptibility of the Polyakov loop $\chi_P$ is measured to precisely identify the point where the center symmetry is spontaneously broken. The peak of $\chi_P$ is a clear signal of the deconfinement phase transition, see for example Fig.~\ref{SusceptibilityPolyakov12_1650b.pdf}.

\begin{figure}[t]
\centering
\subfigure[Finite size scaling]{\includegraphics[width=0.43\textwidth]{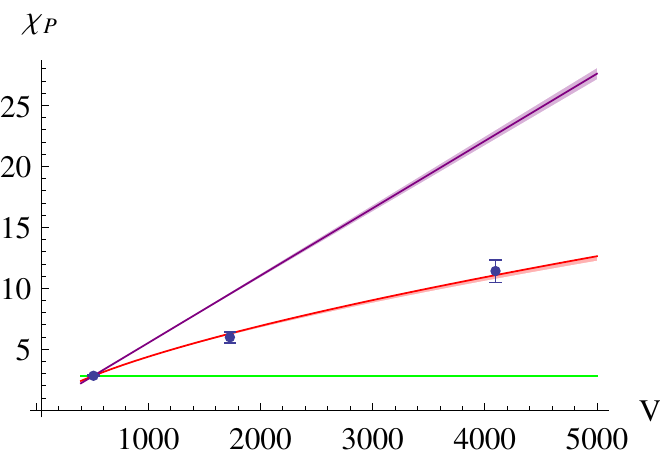}\label{finitesizescaling_polyakovloop_1650b}}
\subfigure[]{\includegraphics[width=0.47\textwidth]{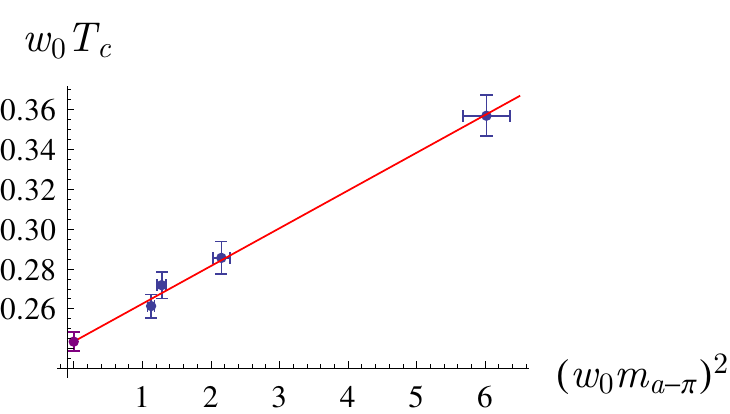}\label{finite_temperature_chiral_extrapolation_mass_independent}}
\caption{a) Scaling of the value of the susceptibility of the Polyakov loop at the peak for a lattice $N_s^3 \times N_t$, with $N_t = 4$ and $N_s = \{8,12,16\}$, $\beta=1.65$ and $\kappa=0.16$. The colored lines are obtained from formula ({\protect \ref{scaling}}), using $V_1 = 8^3$ for the extrapolation to the other volumes $V_2=12^3$ and $V_2 = 16^3$. The purple line marks the expected scaling for a first order phase transition, the red line a second order phase transition with $x = 0.654$ and the green line a crossover. b) Extrapolation to the chiral limit of the critical deconfinement temperature.}
\end{figure}

The order of the phase transition can be estimated from the maximal value of the susceptibility  $\chi_{\textrm{max}}(V)$ considered as a function of the spatial volume $V$. The exponent $x$ in the expected scaling
\begin{equation}\label{scaling}
 \frac{\chi_{\textrm{max}}(V_2)}{\chi_{\textrm{max}}(V_1)} = \left(\frac{V_2}{V_1}\right)^x
\end{equation}
is equal to 1 for a first order phase transition, to 0 for a crossover and equal to $0.654(7)$ for a second order phase transition in the universality class of the three dimensional $Z_2$ spin Ising model \cite{Ferrenberg1991}. As shown in Fig.~\ref{finitesizescaling_polyakovloop_1650b}, our results are in good agreement with a second order phase transition, meaning that the order of the phase transition is unchanged with respect to pure Yang-Mills theory by the gluino interactions in the region explored. In particular, the Svetitsky-Yaffe conjecture \cite{Svetitsky:1982gs} seems to hold also in our model.

The peak of the susceptibility in Fig.~\ref{SusceptibilityPolyakov12_1650b.pdf} identifies the deconfinement phase transition for a given $\mathcal{N}=1$ SYM theory softly broken by a specific bare gluino mass $m_0 \simeq -0.875$. Many different simulations have been done for different lattice sizes and different gauge couplings. The peak of $\chi_P$ defines a critical line of couplings $(\beta_c,\kappa_c)$ in the bare parameter space. This line can be used to extrapolate the critical temperature to the supersymmetric limit. The critical temperature is defined as
\begin{equation}
 T_c = \frac{1}{N_t a(\beta_c)}\,,
\end{equation}
and the lattice spacing $a(\beta)$ is set at zero temperature in terms of the scale $w_0$ in a mass independent renormalization scheme \cite{Borsanyi:2012zs}. A proper definition of the gluino mass is required to extrapolate $T_c$ to the chiral limit, i.e. to the point where the renormalized gluino mass $m_R$ vanishes. The adjoint pion mass $m_{a-\pi}$, defined in a partially quenched approach, is proportional to the square root of $m_R$, and it provides a precise definition of the chiral limit \cite{Munster:2014cja}. We set therefore $m_R = (w_0 m_{a-\pi})^2$, up to an inessential multiplicative factor.

The final extrapolation is presented in Fig.~\ref{finite_temperature_chiral_extrapolation_mass_independent}. The critical temperature decreases approaching the chiral limit. The linear fit
\begin{equation}
 (w_0 T_c)(m_R) =   0.0190(22) m_R + 0.2432(45)\,,
\end{equation}
evaluated at $m_R = 0$, defines the critical temperature for the $\mathcal{N}=1$ super Yang-Mill theory
\begin{equation}
 w_0 T_c|_{m_R = 0} = 0.2432(45)\,,
\end{equation}
up to lattice discretization errors. The lattice action has been sampled also using one-level stout smeared links in the definition of the Dirac-Wilson operator, in order to roughly confirm the main conclusions obtained. No differences have been found in the qualitative behavior; the final result
\begin{equation}
w_0 T_c|_{m_R = 0} =  0.242(37)
\end{equation}
is compatible with the above extrapolation. We have computed the critical temperature for pure SU(2) gauge theory in terms of $w_0$, to have a comparison and an estimation of the effects of the gluino interactions on $T_c$. The result
\begin{equation}
 w_0 T_c|_{m_R = \infty} =  0.2941(13)
\end{equation}
leads to the ratio
\begin{equation}
 \frac{T_c(\textrm{SYM})}{T_c(\textrm{pure Yang-Mills})} = 0.826(18)\,,
\end{equation}
dependent in principle on the common observable chosen to set the scale (in this case $w_0$).

\section{Chiral symmetry restoration}

\begin{figure}[tb]
\centering
\includegraphics[width=0.41\textwidth]{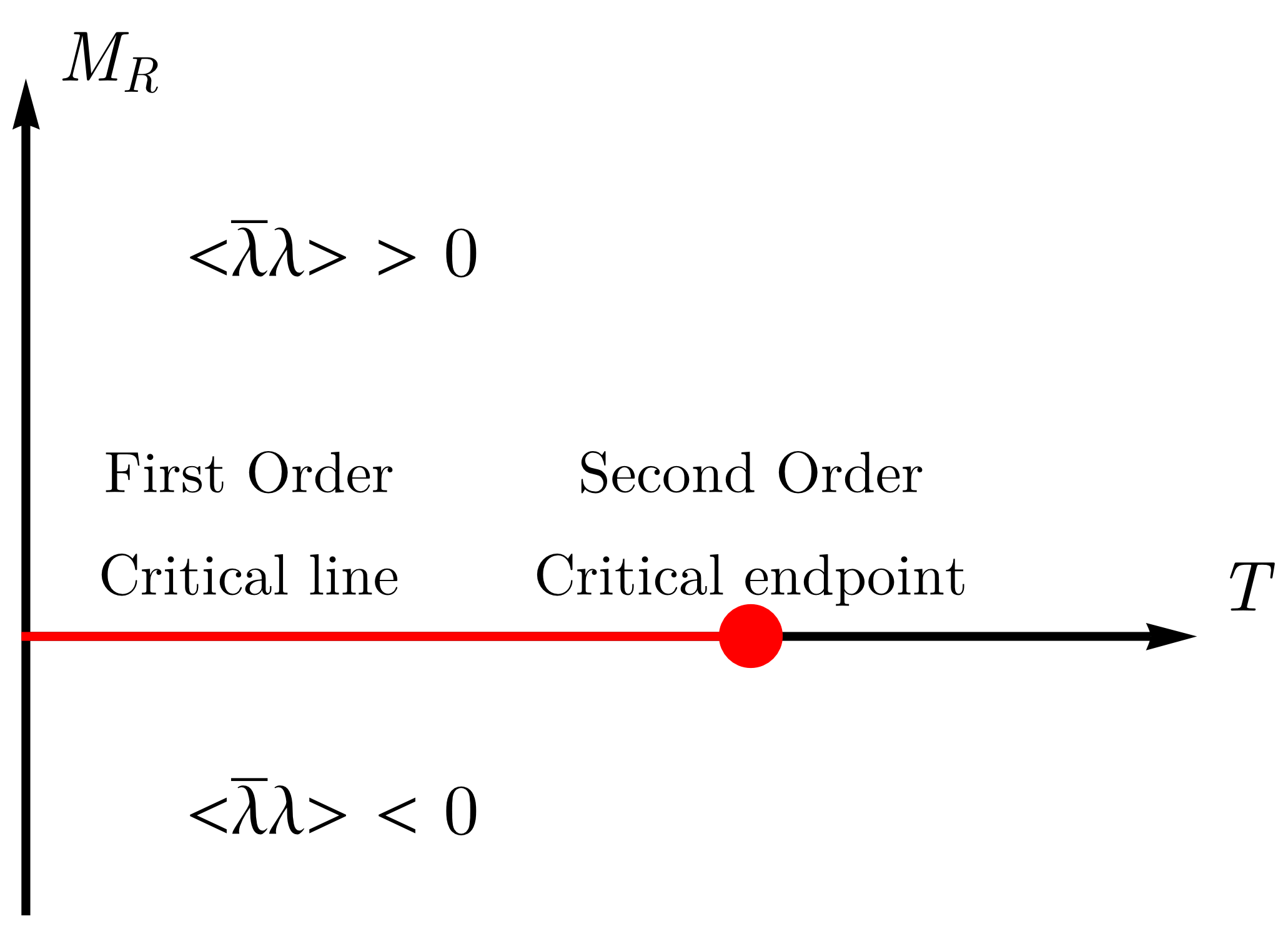}
\caption{Expected phase diagram for chiral symmetry breaking of $\mathcal{N}=1$ SYM with two colors.}\label{phase_chiral}
\end{figure}

Chiral symmetry is generated by the U$_A$(1) group of transformations
\begin{equation}
 \lambda \rightarrow \exp\{\I\alpha \gamma_5 \} \lambda\,,
\end{equation}
and it is broken down to the discrete subgroup $Z_{2 N_c}$ due to the axial anomaly. The group $Z_{2 N_c}$ is further spontaneously broken down to $Z_2$ by a non-vanishing expectation value of the gluino condensate $\langle \bar{\lambda}\lambda\rangle$, the full pattern is
\begin{equation}
 \textrm{U}_A(1) \underset{\textrm{anomaly}}{\xrightarrow{\hspace*{1.3cm}}} Z_{2N_c} \underset{\langle \bar{\lambda}\lambda\rangle \neq 0}{\xrightarrow{\hspace*{1.3cm}}} Z_2\,.
\end{equation}
At zero temperature, the gluino condensate jumps with a first order phase transition crossing the critical line defined by the condition $m_R = 0$ \cite{Amati:1988ft,Seiberg:1994pq,Kirchner:1998mp}. The gluino mass plays therefore a similar role to the external magnetic field $B$ in the Ising model, in particular the gluino condensate is in correspondence with the spontaneous magnetization. If the analogy with the Ising model is correct, a second order phase transition is expected in the supersymmetric theory for the gauge group SU(2), occurring at a critical temperature $T_c^{A}$ where the $Z_{4}$ chiral symmetry is restored, see Fig.~\ref{phase_chiral}.

There are in principle no known relations between $T_c^{A}$ and the deconfinement temperature $T_c^{D}$ discussed in the section before. An interesting possibility arises when both phase transitions coincide, but it could be as well that chiral and center symmetry are broken or restored at two different critical temperatures.

We have done some preliminary simulations to address this question. The bare chiral condensate is simply the expectation value of the trace of the inverse of the Dirac-Wilson operator,
but it requires an additive and multiplicative renormalization. The subtracted chiral condensate
is computed to get rid of the additive renormalization constant, while we avoid the calculation of the multiplicative renormalization factor by fixing the gauge coupling to $\beta = 1.7$. The temperature is changed by varying $N_t$. The lattice chosen has a volume $12^3\times N_t$, with $N_t$ in the range 4--12. The bare gluino mass is fixed to $\kappa = 0.192$; the adjoint pion mass is $am_{a-\pi}=0.388(9)$ and the value of the scale is $w_0/a=2.070(38)$.

\begin{figure}[tb]
\centering
\subfigure[Chiral susceptibility]
{\includegraphics[width=0.45\textwidth]{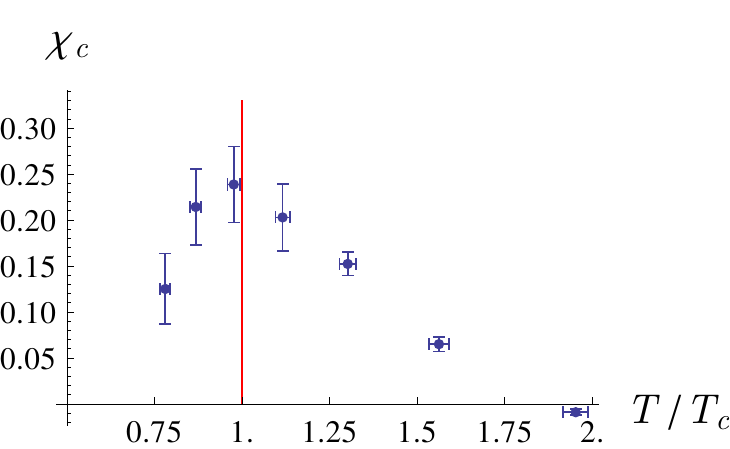}
\label{chiralcondsusc12c}}
\subfigure[Polyakov loop susceptibility]
{\includegraphics[width=0.45\textwidth]{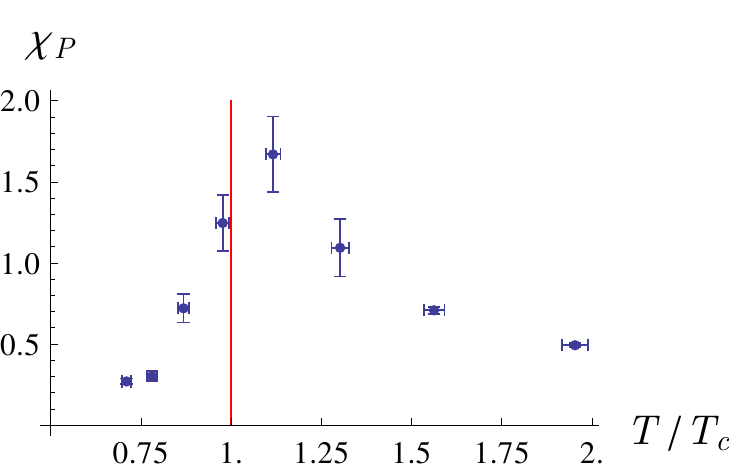}
\label{polyakovsusc12c}}
\caption{Lattice $12^3\times N_t$, $\beta = 1.7$ and $\kappa = 0.192$; a) susceptibility of the chiral condensate and b) Polyakov loop susceptibility. Note that the two peaks roughly coincide. The red line marks the deconfinement temperature obtained from the  simulations discussed in the previous section.}
\end{figure}

We have computed the chiral susceptibility $\chi_c$ to identify as before the critical point by its peak, see Fig.~\ref{chiralcondsusc12c}. A direct comparison with the Polyakov loop susceptibility seems to support the possibility that chiral symmetry restoration and deconfinement occur at about the same temperature, see Fig.~\ref{polyakovsusc12c}. We have done some further simulations closer to the critical point and we have been able to see no evidence for chiral symmetry breaking up to a temperature higher than approximately twice the deconfinement one. 

\section{Conclusions}

We have presented the first numerical investigation of the phase diagram of the $\mathcal{N} = 1$ super Yang-Mills theory at finite temperatures. We have found that the deconfinement phase transition has a good signal for all values of the gluino mass and that the finite volume scaling seems to be in agreement with a second order phase transition in the universality class of the three dimensional $Z_2$ Ising model.

Our preliminary investigations support the possibility that chiral restoration occurs roughly at the same temperature of the deconfinement phase transition. {}From this perspective $\mathcal{N} = 1$ SYM appears to be similar to QCD. Further studies closer to the continuum and to the massless limit are required to fully support this scenario.

A possible direction of future research is the study of $\mathcal{N}=1$ SYM compactified with periodic boundary conditions applied to fermions; first investigations have been presented in Ref.~\cite{Bergner2014compact}.

\section*{Acknowledgments}

The authors thankfully acknowledge the computing time granted by the John von Neumann Institute for Computing (NIC) and provided on the supercomputer JUROPA at J\"ulich Supercomputing Centre (JSC). Additional computational resources have been provided by the computer cluster PALMA of the University of M\"unster.

\end{document}